\documentclass[prb,twocolumn,showpacs,preprintnumbers,amsmath,amssymb,superscriptaddress]{revtex4}
\usepackage{graphicx}
\usepackage{amssymb}
\usepackage{natbib}
\begin{document}
\title{Polarized neutron scattering studies of magnetic excitations in
electron-overdoped superconducting BaFe$_{1.85}$Ni$_{0.15}$As$_{2}$}
\author{Mengshu Liu}
\affiliation{
Department of Physics and Astronomy, The University of Tennessee, Knoxville, Tennessee 37996-1200, USA
}
\author{C. Lester}
\affiliation{
H. H. Wills Physics Laboratory, University of Bristol, Tyndall Avenue,      
Bristol, BS8 1TL, UK 
}
\author{Jiri Kulda}
\affiliation{
Institut Laue-Langevin, 6, rue Jules Horowitz, BP 156, 38042 Grenoble Cedex 9, France
}
\author{Xinye Lu}
\affiliation{Beijing National Laboratory for Condensed Matter Physics,
Institute of Physics, Chinese Academy of Sciences, Beijing 100190, China
}
\affiliation{
Department of Physics and Astronomy, The University of Tennessee, Knoxville, Tennessee 37996-1200, USA
}
\author{Huiqian Luo}
\affiliation{Beijing National Laboratory for Condensed Matter Physics,
Institute of Physics, Chinese Academy of Sciences, Beijing 100190, China
}
\author{Meng Wang}
\affiliation{Beijing National Laboratory for Condensed Matter Physics,
Institute of Physics, Chinese Academy of Sciences, Beijing 100190, China
}
\author{Stephen M. Hayden}
\affiliation{
H. H. Wills Physics Laboratory, University of Bristol, Tyndall Avenue,      
Bristol, BS8 1TL, UK 
}
\author{Pengcheng Dai}
\affiliation{
Department of Physics and Astronomy, The University of Tennessee, Knoxville, Tennessee 37996-1200, USA
}
\affiliation{Beijing National Laboratory for Condensed Matter Physics,
Institute of Physics, Chinese Academy of Sciences, Beijing 100190, China
}
\date{\today}
\begin{abstract}
We use polarized inelastic neutron scattering to study low-energy
spin excitations and their spatial anisotropy in
electron-overdoped superconducting
BaFe$_{1.85}$Ni$_{0.15}$As$_{2}$ ($T_c=14$ K). In the normal
state, the imaginary part of the dynamic susceptibility,
$\chi^{\prime\prime}(Q,\omega)$, at the antiferromagnetic (AF)
wave vector $Q=(0.5,0.5,1)$ increases linearly with energy for
$E\le 13$ meV. Upon entering the superconducting state, a spin gap
opens below $E\approx 3$ meV and a broad neutron spin resonance
appears at $E\approx 7$ meV.  Our careful neutron polarization
analysis reveals that $\chi^{\prime\prime}(Q,\omega)$ is isotropic
for the in-plane and out-of-plane components in both the normal
and superconducting states. A comparison of these results with
those of undoped BaFe$_2$As$_2$ and optimally electron-doped
BaFe$_{1.9}$Ni$_{0.1}$As$_{2}$ ($T_c=20$ K) suggests that the spin
anisotropy observed in BaFe$_{1.9}$Ni$_{0.1}$As$_{2}$ is likely
due to its proximity to the undoped BaFe$_2$As$_2$. Therefore, the
neutron spin resonance is isotropic in the overdoped regime,
consistent with a singlet to triplet excitation.
\end{abstract}
\maketitle
\section{Introduction}

Understanding the role of spin excitations in the
superconductivity of iron arsenides \cite{Kamihara,Rotter,ljli} is
important for developing a microscopic theory of superconductivity
in these materials \cite{mazin,korshunov,maier1,maier2,seo2}. Like
copper oxide superconductors, superconductivity in iron pnictides
arises when electrons or holes are doped into their
antiferromagnetically-ordered parent compounds \cite{Cruz}. For
electron-doped BaFe$_{2-x}T_x$As$_2$ ($T=$Co, Ni) \cite{ljli}, the
antiferromagnetic (AF) and superconducting phase diagrams as a
function of Co(Ni)-doping have been determined by neutron
scattering experiments (Fig. 1(a)) \cite{clester09,hqluo}. Near
the optimally electron-doped superconductor
BaFe$_{2-x}$Ni$_x$As$_2$ at $x=0.1$ ($T_c=20$ K), the static AF
order is suppressed \cite{chi}. However, short-range spin
excitations persist and couple directly to superconductivity via a
collective magnetic excitation termed the neutron spin resonance
\cite{chi,lumsden,slli,inosov,mywang,jtpark}.  As a function of
Ni-doping, the energy of the resonance is associated with both the
superconducting electronic gap $\Delta$ and $k_BT_c$, thus
indicating its direct connection with superconductivity
\cite{dsinosov}.

Although the resonance appears to be a common feature amongst
different classes of unconventional superconductors including
high-$T_c$ copper oxides \cite{mignod,mook,fong,dai,wilson}, heavy
Fermions \cite{metoki,stock}, and iron-based materials
\cite{chi,lumsden,slli,inosov,mywang,jtpark,mook10,qiu,harriger10},
much remains unknown about its microscopic origin.  Assuming that
the resonance is a spin-1 singlet-to-triplet excitation of the
Cooper pairs \cite{eschrig}, it should be possible to split it
into three peaks under the influence of a magnetic field via the
Zeeman effect by an amount $\Delta E=\pm g\mu_B B$, where $g = 2$
is the Lande factor and $B$ is the magnitude of the field
\cite{dai00,jwen10,slli11,stock12}. Although there have been
attempts to split the resonance for copper oxide \cite{dai00} and
iron-based superconductors \cite{jwen10,slli11} in this way, the
results are inconclusive and it has not been possible determine if
the mode is indeed a singlet-to-triplet excitation. In a very
recent neutron experiment performed on the heavy Fermion
superconductor CeCoIn$_5$, the resonance was shown to be a doublet
excitation \cite{stock12}, thus casting doubt on its direct
connection with superconducting Cooper pairs \cite{jzhao10}.

Alternatively, one can use neutron polarization analysis to
determine the nature of the resonance and the effect of
superconductivity on spin excitations. If the resonance is an
isotropic triplet excitation of the singlet superconducting ground
state, one expects that the degenerate triplet would be isotropic
in space. Utilising neutron polarization analysis, one can
conclusively separate the magnetic signal from lattice scattering
and determine the spatial anisotropy of the magnetic excitations
\cite{moon}. For the optimally hole-doped copper oxide
superconductor YBa$_2$Cu$_3$O$_{6.9}$ \cite{mignod,mook,fong,dai},
recent polarized neutron scattering experiments reveal that while
the resonance at $E=41$ meV is isotropic in space, magnetic
excitations below the resonance ($10\leq E\leq 30$ meV) exhibit
large anisotropy with the excitations polarized along the $c$-axis
being suppressed \cite{headings}.  These results suggest that
while the resonance itself is consistent with a spin-1
singlet-to-triplet excitation, the emergence of low-energy spin
anisotropy may arise from the spin-orbit coupling due to the
Dzyaloshinskii-Moriya interactions between the copper spins
\cite{coffey}. In the case of iron-based superconductors, the
situation is more complicated. For optimally electron-doped
BaFe$_{1.9}$Ni$_{0.1}$As$_2$, polarized neutron scattering
experiments indicate that while the magnetic scattering is
essentially isotropic in the normal state, a large spin anisotropy
develops below $T_c$. Excitations polarized along the $c$-axis are
larger than those in the plane for energies $2\leq E\leq 6$ meV,
i.e. below the weakly anisotropic resonance \cite{lipscombe10}. On
the other hand, similar measurements on superconducting
FeSe$_{0.5}$Te$_{0.5}$ reveal an anisotropic resonance with the
in-plane component slightly larger than the out-of-plane component
\cite{babkevich}.  However, the spin excitations are isotropic for
energies below and above the resonance \cite{babkevich}. Finally,
recent neutron polarization analysis of spin waves in the undoped
AF BaFe$_2$As$_2$ \cite{qureshi12} indicate that the magnetic
single-ion anisotropy induced spin-wave gaps \cite{jzhao08,matan}
are strongly anisotropic, with the in-plane component of the
spin-wave gap much larger than that of the $c$-axis component.
Therefore, it costs more energy to rotate a spin within the
orthorhombic $a$-$b$ plane than rotating it perpendicular to the
FeAs layers in the AF ordered state of BaFe$_2$As$_2$
\cite{qureshi12}.

\begin{figure}
\includegraphics[scale=.35]{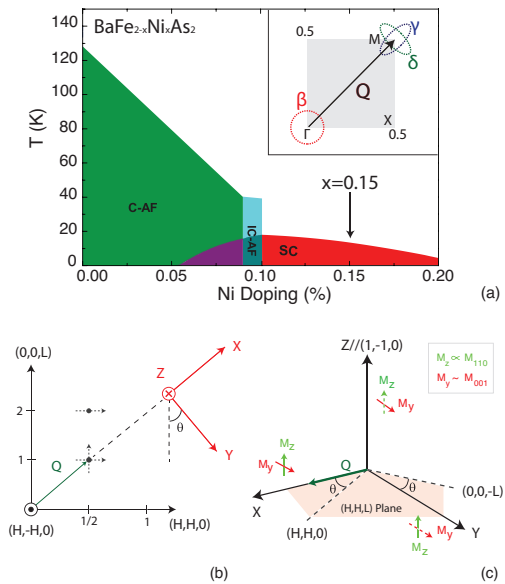}
\caption { (Color online) (a) The schematic antiferromagnetic and
superconducting phase diagram of BaFe$_{2-x}$Ni$_x$As$_2$ as
determined from neutron diffraction experiments \cite{hqluo}.  The
present composition is highlighted with an arrow. The inset shows
an illustration of quasiparticle excitations from the hole Fermi
pocket at the $\Gamma$ point to the electron pocket at the $M$
point as predicted by Fermi surface nesting theories. (b) The
three neutron polarization directions ($x$, $y$ and $z$) oriented
in  the $(H,H,L)$ plane of the reciprocal space. (c) The
relationship between magnetic components $M_y$ and $M_z$ measured
by polarized neutron scattering and in-plane ($M_{110}$) and
out-of-plane ($M_{001}$) dynamic spin susceptibility. The solid
arrow denotes the measured magnetic component in a SF channel and
the dashed arrow denotes the component measured in a NSF channel.
In this geometry, we have $M_z \propto M_{1\bar{1}0}=M_{110}$, due
to tetragonal symmetry; and $M_y \sim M_{001}$, given that
$\theta$ is a small value. }
\end{figure}

Given the current confusing experimental situation on the
anisotropy of spin excitations in undoped and optimally
electron-doped BaFe$_{2-x}$Ni$_x$As$_2$
\cite{lipscombe10,qureshi12}, it would be interesting to carry out
similar polarized neutron scattering measurements for electron
overdoped BaFe$_{2-x}$Ni$_x$As$_2$.  From the electronic phase
diagram of BaFe$_{2-x}$Ni$_x$As$_2$ in Fig. 1(a) \cite{hqluo}, we
see that samples in the overdoped regime are far from the AF and
superconductivity co-existence region, and thus avoid possible
influence of the local magnetic anisotropy present in undoped
BaFe$_2$As$_2$ \cite{qureshi12}. For our neutron experiments, we
prepared over-doped BaFe$_{1.85}$Ni$_{0.15}$As$_{2}$ with $T_c=14$
K (Fig. 1(a)). In this article, we describe polarized neutron
scattering studies of energy and momentum dependence of the
magnetic excitations in BaFe$_{1.85}$Ni$_{0.15}$As$_{2}$ below and
above $T_c$. We find that the spin excitations at or near the
resonance energy are spatially isotropic. By comparing these
results with previous work on undoped BaFe$_2$As$_2$ and optimally
doped BaFe$_{1.9}$Ni$_{0.1}$As$_2$ \cite{lipscombe10,qureshi12},
we conclude that the strong in-plane single-ion anisotropy in
antiferromagnetically-ordered orthorhombic BaFe$_2$As$_2$ extends
to the paramagnetic tetragonal BaFe$_{1.9}$Ni$_{0.1}$As$_2$,
giving rise to the large out-of-plane component of the low-energy
spin excitations for the superconducting
BaFe$_{1.9}$Ni$_{0.1}$As$_2$.  Therefore, the resonance in
optimally and overdoped  BaFe$_{2-x}$Ni$_x$As$_2$ ($x=0.1,0.15$)
is mostly isotropic in space, consistent with the
singlet-to-triplet excitation scenario.

\begin{figure}
\includegraphics[scale=.5]{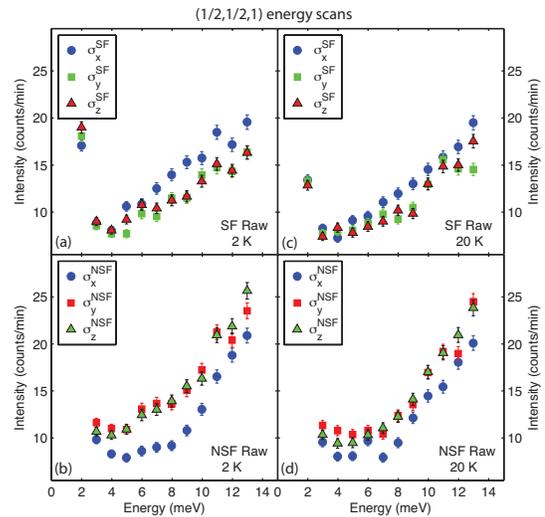}
\caption { (Color online)  Constant-$Q$ scans at ${\bf
Q}=(0.5,0.5,1)$ below and above $T_c$. Using polarized neutrons,
we can measure six independent scattering cross sections: incoming
neutrons polarized along the $x$, $y$ or $z$ directions, with
outgoing neutrons flipped (SF), or not flipped (NSF). (a) The raw
data for SF scattering at 2 K, denoted as
$\sigma_{x,y,z}^{\textrm{SF}}$; (b) Identical scans in NSF
channel, or $\sigma_{x,y,z}^{\textrm{NSF}}$; (c) SF scattering
$\sigma_{x,y,z}^{\textrm{SF}}$ at 20 K, and (d) NSF scattering
$\sigma_{x,y,z}^{\textrm{NSF}}$ at 20 K. }
\end{figure}

\section{Experimental details}

We grew large single crystals of the overdoped iron arsenide
superconductor BaFe$_{1.85}$Ni$_{0.15}$As$_{2}$ using a self-flux
method \cite{ycchen}. BaFe$_{1.85}$Ni$_{0.15}$As$_{2}$ has a $T_c$
of $14$ K, and is far away from the AF phase of the undoped
BaFe$_2$As$_2$ (Fig.1(a)).  As a function of increasing Ni-doping
$x$, the low-temperature crystal structure of
BaFe$_{2-x}$Ni$_x$As$_2$ changes from orthorhombic to tetragonal
with $a=b$ for $x\ge 0.1$ \cite{hqluo,chi}. For this experiment,
we coaligned $\sim 15$ g single crystals of
BaFe$_{1.85}$Ni$_{0.15}$As$_{2}$ in the $(H,H,L)$ scattering plane
(with mosaicity $3^\circ$ at full width half maximum) with a
tetragonal unit cell for which $a=b=3.96$ \AA, and $c=12.77$ \AA.
In this notation, the vector \textbf{Q} in three-dimensional
reciprocal space in \AA$^{-1}$ is defined as
$\textbf{Q}=H\textbf{a} ^*+K\textbf{b} ^*+L\textbf{c} ^*$, where
$H$, $K$, and $L$ are Miller indices and $\textbf{a}
^*=\hat{\textbf{a}}2\pi/a, \textbf{b}
^*=\hat{\textbf{b}}2\pi/b,\textbf{c} ^*=\hat{\textbf{c}}2\pi/c$
are reciprocal lattice vectors.

\begin{figure}
\includegraphics[scale=.5]{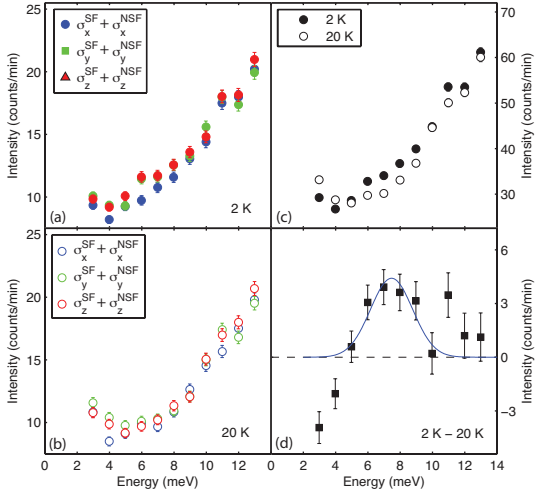}
\caption { (Color online) (a) Simulation of unpolarized energy
scans using
$\sigma_{\alpha}^{\textrm{SF}}+\sigma_{\alpha}^{\textrm{NSF}}$
with $\alpha=x,y,z$ at 2 K and (b) 20 K.  The wave vector is fixed
at ${\bf Q}=(0.5,0.5,1)$. (c) Unpolarized energy scan at
$(1/2,1/2,1)$ below and above $T_c$ obtained by adding all six
channels together. (d) Temperature difference plot between 2 K and
20 K reveals a neutron spin resonance at $E =7$ meV and negative
scattering below 4 meV, very similar to the earlier unpolarized
measurements on the same Ni-doping level \cite{mywang}. }
\end{figure}

\begin{figure}
\includegraphics[scale=.5]{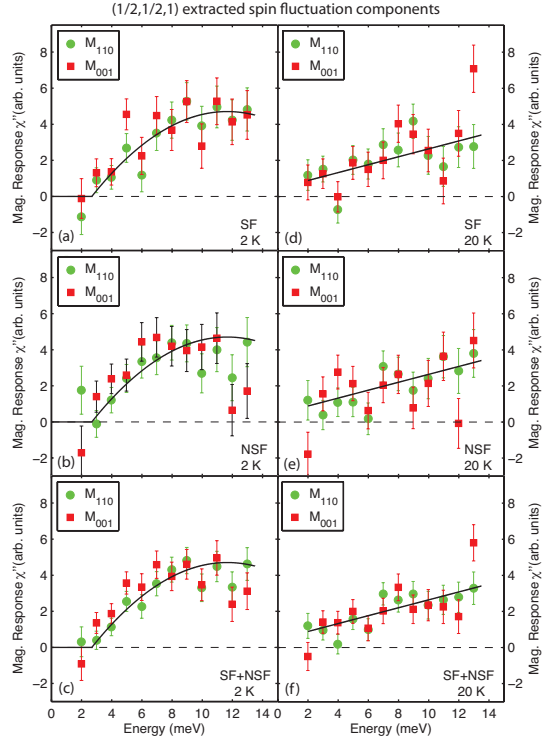}
\caption { (Color online) Neutron polarization analysis used to
extract the in-plane ($M_{110}$) and out-of-plane $M_{001}$
components of spin excitations in BaFe$_{1.85}$Ni$_{0.15}$As$_2$
from SF and NSF data in Fig. 2.  $M_{110}$ and $M_{001}$ at 2 K
are extracted from (a) SF, and (b) NSF data in Fig.2. (d,e)
$M_{110}$ and $M_{001}$ at 20 K. The above analysis is based 
on the assumption that the background scattering for the $x, y,$ 
and $z$ spin polarizations are different (see eqs. (2) and (3)). 
However, if we assume that backgrounds are identical for different spin 
polarizations, we would obtain higher magnetic scattering intensity 
in the NSF channel compared with that of the SF channel at all 
measured temperatures and energies. At present, the microscopic 
origin of such a difference is unclear. (c) The combination of SF and NSF
data at 2K. (f) The combination of SF and NSF data at 20K. These
data reveal isotropic paramagnetic scattering at the probed
energies and temperatures. }
\end{figure}

We carried out polarized inelastic neutron scattering experiments
at the IN20 triple-axis spectrometer at the Institut Laue-Langevin
in Grenoble, France.  We used the Cryopad capability of the IN20
spectrometer in order to ensure that the sample was in a strictly
zero magnetic field environment. This avoids errors due to flux
inclusion and field expulsion in the superconducting phase of the
sample. Polarized neutrons were produced using a focusing Heusler
monochromator and analyzed using a focusing Heusler analyzer 
with a fixed final wave vector at $k_f=2.662$\AA$^{-1}$. 
To facilitate easy comparison with previous polarized neutron
scattering results \cite{lipscombe10}, we define neutron
polarization directions as $x,y,z$, with $x$ parallel to
\textbf{Q} and $y$ and $z$ both perpendicular to \textbf{Q} as
shown in Fig. 1(b). Since neutron scattering is only sensitive to
those magnetic scattering components perpendicular to the momentum
transfer \textbf{Q}, magnetic responses within the $y-z$ plane
($M_y$ and $M_z$) can be measured. At a specific momentum and
energy transfer, incident neutrons can be polarized along the $x$,
$y$, and $z$ directions, and the scattered neutrons can have
polarizations either parallel (neutron nonspin flip or NSF,
$\uparrow\uparrow$) or antiparallel (neutron spin flip or SF,
$\uparrow\downarrow$) to the incident neutrons. Therefore, the six
neutron scattering cross sections can be written as
$\sigma_{\alpha}^{\rm NSF}$ and $\sigma_{\alpha}^{\rm SF}$, where
$\alpha=x,y,z$ \cite{moon,lipscombe10}.  If we use $M_\alpha$ and
$N$ to denote the magnetic response and nuclear scattering,
respectively, the neutron scattering cross sections
$\sigma_{\alpha}^{\rm NSF}$ and $\sigma_{\alpha}^{\rm SF}$ are
related to $M_\alpha$ and $N$ via Eq. (1):
\begin{equation}
\left( \begin{array}{ccc}
\sigma_x^{\textrm{SF}} \\
\sigma_y^{\textrm{SF}} \\
\sigma_z^{\textrm{SF}} \\
\sigma_x^{\textrm{NSF}} \\
\sigma_y^{\textrm{NSF}} \\
\sigma_z^{\textrm{NSF}}
\end{array} \right)\ = \left( \begin{array}{ccc}
1 & 1 & 0 \\
0 & 1 & 0 \\
1 & 0 & 0 \\
0 & 0 & 1 \\
1 & 0 & 1 \\
0 & 1 & 1
\end{array} \right)\ \times \left( \begin{array}{ccc}
M_y \\
M_z \\
N
\end{array} \right)
\end{equation}

 \begin{figure}
\includegraphics[scale=.5]{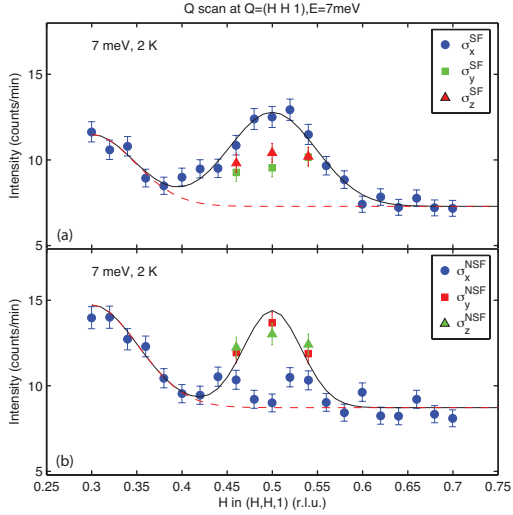}
\caption { (Color online) Constant-energy scans along the
$[H,H,1]$ direction at the resonance energy of $E=7$ meV at 2 K
for different neutron polarization directions. (a) Neutron SF
scattering cross sections for the $x$, $y$, and $z$ polarization
directions. (b) NSF scattering cross sections.  A clear peak is
seen at ${\bf Q}=(0.5,0.5,1)$ in the $\sigma_{x}^{\rm SF}$ channel
that is absent in the  $\sigma_{x}^{\rm NSF}$ channel, thus
confirming the magnetic nature of the resonance. }
\end{figure}

In a real experiment, neutron polarization is not 100\% and there
are also neutron spin independent backgrounds (nuclear-spin
incoherent scattering and general instrumental background). Since
neutron SF and NSF scattering processes have identical
instrumental setups and only the spin directions of the incident
neutrons are changed, we assume constant backgrounds of $B_1$,
$B_2$, $B_3$ for neutron polarizations in the $x$, $y$, and $z$
directions, respectively. We have measured the neutron flipping
ratios $R$ for all three neutron polarizations, and found them to
be independent of neutron polarization directions within the
errors of our measurements.  By considering finite flipping ratio
and assume that instrumental backgrounds for different neutron 
polarizations are slightly different, we have
\begin{equation}
\left( \begin{array}{ccc}
\sigma_x^{\textrm{SF}}-B_1 \\
\sigma_y^{\textrm{SF}}-B_2 \\
\sigma_z^{\textrm{SF}}-B_3 \\
\sigma_x^{\textrm{NSF}}-B_1 \\
\sigma_y^{\textrm{NSF}}-B_2 \\
\sigma_z^{\textrm{NSF}}-B_3
\end{array} \right)\ = \frac{1}{R+1} \left( \begin{array}{ccc}
R & R & 1 \\
1 & R & 1 \\
R & 1 & 1 \\
1 & 1 & R \\
R & 1 & R \\
1 & R & R
\end{array} \right)\ \times \left( \begin{array}{ccc}
M_y \\
M_z \\
N
\end{array} \right),
\end{equation}
where the flipping ratio $R$ is measured by the leakage of NSF
nuclear Bragg peaks into the magnetic SF channel
$R=\sigma_{Bragg}^\textrm{NSF}/\sigma_{Bragg}^\textrm{SF} \approx
14$. The magnetic moments $M_y$ and $M_z$ can be extracted from
Eq.(2) via
\begin{equation}
\left\{
\begin{array}{ccc}
\sigma_x^{\textrm{SF}}-\sigma_y^{\textrm{SF}}+B_1=\sigma_y^{\textrm{NSF}}-\sigma_x^{\textrm{NSF}}-B_1=cM_y,\\
\\[1pt]
\sigma_x^{\textrm{SF}}-\sigma_z^{\textrm{SF}}+B_2=\sigma_z^{\textrm{NSF}}-\sigma_x^{\textrm{NSF}}-B_2=cM_z
\end{array}
\right.
\end{equation}
where $c=(R-1)/(R+1)$, and $B_1,B_2$ are constant backgrounds. By
measuring all six NSF and SF neutron scattering cross sections, we
can unambiguously determine $M_y$ and $M_z$. To estimate the
in-plane and out-of-plane components of the magnetic scattering
$M_{110}$ and $M_{001}$, we note that $M_{110}=M_{1\bar{1}0}
\equiv M_z$ due to the tetragonal symmetry of the system.
Therefore, $M_{001}$ can be calculated using
$M_y=M_{110}\sin^2\theta+M_{001}\cos^2\theta$. This allows a
complete determination of the temperature and energy dependence of
$M_{110}$ and $M_{001}$.

\begin{figure}
\includegraphics[scale=.5]{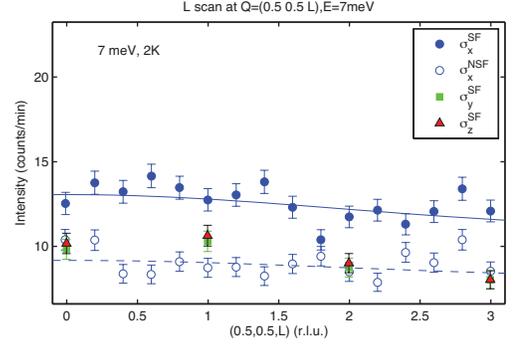}
\caption { (Color online) Constant-energy scans along 
$(0.5,0.5,L)$ at the resonance energy of $E=7$ meV. The
$\sigma_x^{\textrm{SF}}$ and $\sigma_x^{\textrm{NSF}}$ data show
no $L$ dependence. The solid and dashed lines show the expected
magnetic scattering intensity assuming an Fe$^{2+}$ form factor. }
\end{figure}

\section{Results}
In previous polarized neutron scattering experiments performed on
optimally doped BaFe$_{1.9}$Ni$_{0.1}$As$_{2}$ \cite{lipscombe10},
the in-plane ($M_{110}$) and out-of-plane ($M_{001}$) magnetic
fluctuations are gapless and approximately isotropic in the normal
state above $T_c$.   Upon entering the superconducting state, the
$M_{110}$ spectra re-arrange with a spin gap below $E=2$ meV and a
resonance peak near $E=7$ meV.  On the other hand, the $M_{001}$
spectra peak near $E=4$ meV and have a smaller spin gap (Fig. 3 in
Ref. \onlinecite{lipscombe10}). Figures 2(a)-2(d) show all six
constant-${\bf Q}$ scattering cross sections
$\sigma_{x,y,z}^{\textrm{SF}}$ and $\sigma_{x,y,z}^{\textrm{NSF}}$
taken at the AF wave vector $\textbf{Q}=(1/2,1/2,1)$ below and
above $T_c$.  For SF scattering, $\sigma_{y}^{\textrm{SF}}$ is
approximately equal to $\sigma_{z}^{\textrm{SF}}$ at 2 K and 20 K,
but both $\sigma_{y}^{\textrm{SF}}$ and $\sigma_{z}^{\textrm{SF}}$
are smaller than $\sigma_{x}^{\textrm{SF}}$ (Figs. 2(a) and 2(c)).
For the NSF scattering, the situation is similar except that
$\sigma_{x}^{\textrm{NSF}}$ is smaller than
$\sigma_{y}^{\textrm{NSF}}$ and $\sigma_{z}^{\textrm{NSF}}$ (Figs.
2(b) and 2(d)).  These results indicate the presence of
paramagnetic scattering, since for purely nuclear scattering there
would be no difference between the scattering from different
neutron polarizations
($\sigma_{x}^{\textrm{SF}}=\sigma_{y}^{\textrm{SF}}=\sigma_{z}^{\textrm{SF}}$)
\cite{moon}.

\begin{figure}
\includegraphics[scale=.6]{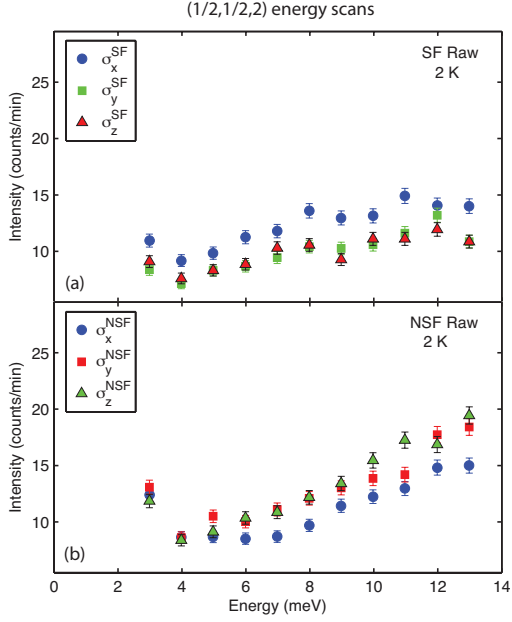}
\caption { (Color online) Constant-$Q$ scans at ${\bf
Q}=(0.5,0.5,2)$ at 2 K. (a) The three neutron SF scattering energy
scans below $T_c$, marked as $\sigma_{x,y,z}^{\textrm{SF}}$. (b)
Identical scans in the neutron NSF channel, marked as
$\sigma_{x,y,z}^{\textrm{NSF}}$. }
\end{figure}

In a previous unpolarized neutron scattering experiment performed
on BaFe$_{1.85}$Ni$_{0.15}$As$_{2}$, a neutron spin resonance was
observed near $E=6$ meV in the superconducting state, found by
taking a temperature difference between constant-${\bf Q}$ scans
at $(0.5,0.5,1)$ r.l.u. \cite{mywang}. Before determining the
possible magnetic anisotropy from neutron polarization analysis,
we note from Eq. (2) that
$\sigma_{x}^{\textrm{SF}}+\sigma_{x}^{\textrm{NSF}}=M_y+M_z+N+2B_1$,
$\sigma_{y}^{\textrm{SF}}+\sigma_{y}^{\textrm{NSF}}=M_y+M_z+N+2B_2$,
and
$\sigma_{z}^{\textrm{SF}}+\sigma_{z}^{\textrm{NSF}}=M_y+M_z+N+2B_3$
are the scattering cross sections for an unpolarized neutron
scattering experiment.  Assuming the background scattering has no
temperature dependence across $T_c$, the temperature difference
data of
$\sigma_{\alpha}^{\textrm{SF}}+\sigma_{\alpha}^{\textrm{NSF}}$
should recover unpolarized neutron scattering results
\cite{mywang}. Figures 3(a) and 3(b) show the sum of the raw data
$\sigma_{\alpha}^{\textrm{SF}}+\sigma_{\alpha}^{\textrm{NSF}}$
above and below $T_c$, respectively for $\alpha=x$, $y$ and $z$.
Figure 3(c) plots the sum of all six scattering cross sections
$\sigma_{x,y,z}^{\textrm{SF}}$ and $\sigma_{x,y,z}^{\textrm{NSF}}$
at $\textbf{Q}=(1/2,1/2,1)$ below and above $T_c$. The temperature
difference in Fig. 3(d) clearly shows a resonant feature at $E=7$
meV, consistent with earlier unpolarized neutron scattering
results \cite{mywang}.

\begin{figure}
\includegraphics[scale=.5]{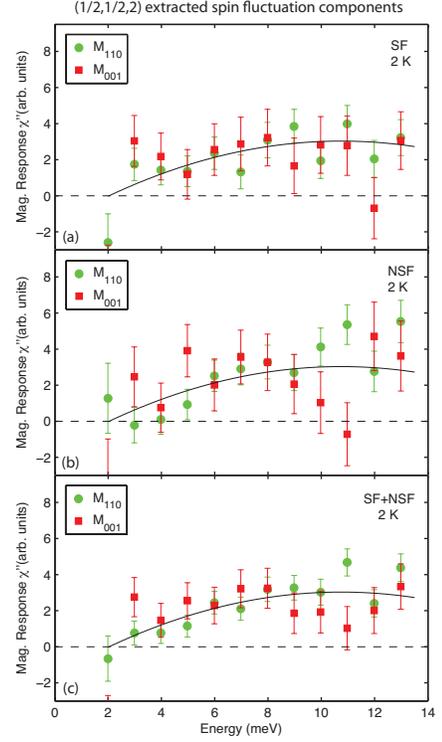}
\caption { (Color online) Constant-$Q$ scans at ${\bf
Q}=(0.5,0.5,2)$ at 2 K. The in-plane ($M_{110}$) and out-of-plane
($M_{001}$) magnetic response extracted from the (a) SF data, and
(b) NSF data, respectively; (c) The combination of SF and NSF data
at 2 K shows no difference between the two magnetic moment
components, indicating isotropic paramagnetic scattering. }
\end{figure}

To extract any possible anisotropy of the resonance and normal
state spin excitations, we use $\sigma_{\alpha}^{\textrm{SF}}$ and
$\sigma_{\alpha}^{\textrm{NSF}}$ with Eq. (3) to independently
determine  $M_y$ and $M_z$.  Since $M_z$ is equal to $M_{110}$ and
$M_y=M_{110}\sin^2\theta+M_{001}\cos^2\theta$, $M_{110}$ and
$M_{001}$ can be independently calculated from either
$\sigma_{\alpha}^{\textrm{SF}}$ or
$\sigma_{\alpha}^{\textrm{NSF}}$.  One can then calculate the
imaginary part of the dynamic susceptibility
$\chi^{\prime\prime}(Q,\omega)$ via
$\chi^{\prime\prime}(Q,\omega)=[1-\exp(-\hbar\omega/k_BT)]S(Q,\omega)$,
where $S(Q,\omega)=M_{110}$ or $M_{001}$, and $E=\hbar\omega$.
Figures 4(a)-4(d) summarize results for
$\chi^{\prime\prime}_{110}(Q,\omega)$ and
$\chi^{\prime\prime}_{001}(Q,\omega)$ at the AF wave vector
$Q=(0.5,0.5,1)$ in the superconducting and normal states,
respectively. The  $\chi^{\prime\prime}_{110}(Q,\omega)$ and
$\chi^{\prime\prime}_{001}(Q,\omega)$ results in Figs. 4(a) and
4(b) are obtained using $\sigma_{\alpha}^{\textrm{SF}}$, while the
similar results shown in Figs. 4(c) and 4(d) are independent
calculations using $\sigma_{\alpha}^{\textrm{NSF}}$.  These
results are identical to within the errors of the measurements.
Figures 4(c) and 4(d) show combined SF+NSF results for
$\chi^{\prime\prime}_{110}(Q,\omega)$ and
$\chi^{\prime\prime}_{001}(Q,\omega)$ to improve the statistics.
In the normal state at 20 K, $\chi^{\prime\prime}_{110}(Q,\omega)$
and $\chi^{\prime\prime}_{001}(Q,\omega)$ are identical and
increase linearly with increasing energy (Fig. 4(f)).  At low
temperatures ($T=2$ K, a spin gap is present below $E\approx 3$
meV and a broad resonance is apparent near $E \approx 7$ meV.
$\chi^{\prime\prime}_{110}(Q,\omega)$ and
$\chi^{\prime\prime}_{001}(Q,\omega)$ are again identical to
within the errors of our measurements.  Therefore, there is no
observable magnetic anisotropy of the spin excitations of
overdoped BaFe$_{1.85}$Ni$_{0.15}$As$_{2}$ in both the normal and
superconducting states at ${\bf Q}=(0.5,0.5,1)$.

Figures 5(a) and 5(b) show constant-energy scans at the resonance
energy along $(H,H,1)$ for
$\sigma_{\alpha}^{\textrm{SF}}$ and
$\sigma_{\alpha}^{\textrm{NSF}}$. While the SF scattering
$\sigma_{x}^{\textrm{SF}}$ shows a clear peak centered at the AF
wave vector ${\bf Q}=(0.5,0.5,1)$ (Fig. 5(a)), the NSF scattering
$\sigma_{x}^{\textrm{NSF}}$ (Fig. 5(b)) is featureless near ${\bf
Q}=(0.5,0.5,1)$. This suggests that the resonance peak above the
background in Fig. 5(a) is entirely magnetic in origin. If the
resonance is purely isotropic paramagnetic scattering, one would
expect $\sigma_{x}^{\textrm{SF}}-B_1\approx
2(\sigma_{y}^{\textrm{SF}}-B_2)\approx
2(\sigma_{z}^{\textrm{SF}}-B_3)$ and
$(\sigma_{y}^{\textrm{NSF}}-B_2)\approx
(\sigma_{z}^{\textrm{NSF}}-B_3)$.  Inspection of Figs. 5(a) and
5(b) reveal that this is indeed the case, thus confirming the
isotropic nature of the magnetic resonance.

To determine whether the spin excitations at the resonance energy
exhibit any $c$-axis modulation in intensity, we carried out
constant-energy scans along $(0.5,0.5,L)$ in the
superconducting state at $E=7$ meV.  As one can see in Fig. 6, the
magnetic scattering intensity decreases smoothly with increasing
$L$, consistent with the expected magnetic intensity reduction due
to the Fe$^{2+}$ form factor (solid line).  There is no evidence
for a $L$-axis modulation of the magnetic scattering.

Finally, to see whether the isotropic magnetic scattering near the
AF wave vector ${\bf Q}=(0.5,0.5,1)$ is independent of the
$c$-axis momentum transfer, we carried out
$\sigma_{\alpha}^{\textrm{SF}}$ and
$\sigma_{\alpha}^{\textrm{NSF}}$ constant-$Q$ scans in the
superconducting state at ${\bf Q}=(0.5,0.5,2)$ (Fig. 7).  Similar
to the data in Fig. 2, the SF scattering
$\sigma_{x}^{\textrm{SF}}$ is larger than
$\sigma_{y}^{\textrm{SF}}$ and $\sigma_{z}^{\textrm{SF}}$ (Fig.
7(a)), while the NSF scattering $\sigma_{x}^{\textrm{NSF}}$ is
smaller than $\sigma_{y}^{\textrm{NSF}}$ and
$\sigma_{z}^{\textrm{NSF}}$. Using this raw data shown in Fig. 7,
we are able to obtain the energy dependence of
$\chi^{\prime\prime}_{110}(Q,\omega)$ and
$\chi^{\prime\prime}_{001}(Q,\omega)$ at ${\bf Q}=(0.5,0.5,2)$ as
shown in Figs. 8(a) and 8(b).  Consistent with the constant-$Q$
scans at ${\bf Q}=(0.5,0.5,1)$, we find isotropic magnetic
scattering at ${\bf Q}=(0.5,0.5,2)$.  Figure 8(c) shows the energy
dependence of $\chi^{\prime\prime}_{110}(Q,\omega)$ and
$\chi^{\prime\prime}_{001}(Q,\omega)$ obtained by combining the SF
and NSF scattering data in Figs. 8(a) and 8(b).  Similar to Fig.
4(c), a spin gap is present below $E=3$ meV and
$\chi^{\prime\prime}_{110}(Q,\omega)$ and
$\chi^{\prime\prime}_{001}(Q,\omega)$ increase with increasing
energy. Therefore, spin excitations in overdoped
BaFe$_{1.85}$Ni$_{0.15}$As$_{2}$ are isotropic below and above
$T_c$ at all energies probed.

\section{Discussion and Conclusions}

In previous polarized neutron scattering experiments on optimally
electron-doped BaFe$_{1.9}$Ni$_{0.15}$As$_{2}$,
$\chi^{\prime\prime}_{110}(Q,\omega)$ and
$\chi^{\prime\prime}_{001}(Q,\omega)$ at  ${\bf Q}=(0.5,0.5,1)$
were found to have peaks near $E=7$ and 4 meV, respectively, in
the superconducting state \cite{lipscombe10}.  These results were
interpreted as being due to the presence of spin-orbital/lattice
coupling \cite{lipscombe10}.  In a recent polarized neutron
scattering work on the AF parent compound BaFe$_{2}$As$_{2}$, it
was found that in-plane polarized magnons exhibit a larger single
iron anisotropy gap than the out-of-plane polarized ones
\cite{qureshi12}. This means that
$\chi^{\prime\prime}_{110}(Q,\omega)$ has a larger gap than
$\chi^{\prime\prime}_{001}(Q,\omega)$ at ${\bf Q}=(0.5,0.5,1)$ in
the AF ordered state, where the Fe moments are locked to the
$a$-axis of the orthorhombic structure
\cite{huang,zhaoprb,goldman} [along the [110] direction in our
tetragonal notation].

From the electronic phase diagram of BaFe$_{2-x}$Ni$_x$As$_2$ in
Fig. 1(a), we see that although the optimally electron-doped
BaFe$_{1.9}$Ni$_{0.1}$As$_{2}$ has tetragonal structure with no
static AF order \cite{chi}, it is very close to that region of the
phase diagram where incommensurate static AF order coexists with
superconductivity \cite{hqluo}.  This suggests that the observed
anisotropy between the in-plane
($\chi^{\prime\prime}_{110}(Q,\omega)$) and out-of-plane
($\chi^{\prime\prime}_{001}(Q,\omega)$) dynamic susceptibility in
tetragonal superconducting BaFe$_{1.9}$Ni$_{0.1}$As$_{2}$
\cite{lipscombe10} may have the same microscopic origin as the
spin wave anisotropy gaps in the AF orthorhombic
BaFe$_{2}$As$_{2}$ \cite{qureshi12}. If this is indeed the case,
the resonance is only weakly anisotropic near optimal
superconductivity, and becomes isotropic in the electron
over-doped BaFe$_{1.9}$Ni$_{0.15}$As$_{2}$.  Therefore, these
results suggest that the resonance in electron over-doped
BaFe$_{2-x}$Ni$_x$As$_2$ is mostly consistent with the
singlet-triplet excitations of electron Cooper pairs.  The
observed spin excitation anisotropy in optimally doped
BaFe$_{2-x}$Ni$_x$As$_2$ is likely due to single iron anisotropy
of spin waves in the parent compound, and suggests that such
anisotropy is present even for samples with tetragonal structure.
Thus, the strong spin-orbital-lattice coupling in electron-doped
BaFe$_{1.9}$Ni$_{0.1}$As$_{2}$ is important for samples up to
optimal superconductivity, and becomes less important for the
overdoped regime.

\section{Acknowledgements}
The work at UTK is supported by the U.S. NSF-DMR-1063866. Work at
IOP is supported by the MOST of China 973 programs (2012CB821400,
2011CBA00110) and NSFC-11004233.

\end{document}